\documentclass[prd,nofootinbib,tightenlines]{revtex4}
\usepackage{graphicx} \usepackage{amsmath} \usepackage{amssymb}
\usepackage{amsfonts} \usepackage{bm}

\usepackage{colordvi}
\usepackage{amsmath,amssymb}
\usepackage{amsfonts}
\usepackage{graphicx,floatflt,amssymb}

\usepackage{amsmath,amssymb}
\usepackage{epsfig}
\usepackage{mathrsfs}
\usepackage{amsmath,amssymb}

\usepackage{amsmath,amssymb, diagrams}

\usepackage{bbm}
\usepackage{bm}
\usepackage{epsfig}
\usepackage{graphicx}

\begin{document}

\newcommand{\be}{\begin{equation}} \newcommand{\ee}{\end{equation}}
\newcommand{\bea}{\begin{eqnarray}}\newcommand{\eea}{\end{eqnarray}}


\title{Twisted Statistics in $\kappa$-Minkowski Spacetime}

\author{T. R. Govindarajan \footnote{trg@imsc.res.in}}

\affiliation{Institute of Mathematical Sciences, CIT Campus, Taramani, Chennai 600113, India}

\author{Kumar S. Gupta \footnote {kumars.gupta@saha.ac.in}}

\affiliation{Theory Division, Saha Institute of Nuclear Physics, 1/AF
Bidhannagar, Calcutta 700064, India}

\author{E. Harikumar \footnote{harisp@uohyd.ernet.in}}

\affiliation{School of Physics, University of Hyderabad, Hyderabad 500046, India}

\author{S. Meljanac \footnote {meljanac@irb.hr} and D. Meljanac
\footnote{dmeljan@irb.hr}}

\affiliation{Rudjer Bo\v{s}kovi\'c Institute, Bijeni\v cka  c.54, HR-10002
Zagreb, Croatia}

\vspace*{1cm}

\begin{abstract}

We consider the issue of statistics for identical particles or fields in $\kappa$-deformed spaces, where the system admits a symmetry group $G$. We obtain the twisted flip operator compatible with the action of the symmetry group, which is relevant for describing particle statistics in presence of the noncommutativity. It is shown that for a special class of realizations, the twisted flip operator is independent of the ordering prescription.
\end{abstract}

\pacs{}

\keywords {$\kappa$ deformed space, noncommutative geometry, statistics }


\maketitle


\section{ Introduction}

Noncommutative geometry is a plausible candidate for describing physics at the Planck scale, a simple model of which is given by the Moyal plane \cite{doug}. The models of noncommutative  spacetime that follow from combining general relativity and uncertainty principle can be much more general \cite{doplicher}. An example of this more general class is provided by the $\kappa$-deformed space \cite{L1,L2,L3}, which is based on a Lie algebra type noncommutativity. Apart from its  algebraic aspects \cite{A2,A3,A4,A5,A7,paper1}, various features of field theories  and symmetries  on $\kappa$-deformed spaces have recently been studied \cite{F1,F2,F3,F4,F5}. Such a space has also been discussed in the context of doubly special relativity \cite{D1,D2,D3}.

The issue of particle statistics plays a central role in the quantum description of a many-body system or field theory. This issue has naturally arisen in the context of noncommutative quantum mechanics and field theory \cite{S1,S2,S3,S4,S6}. In the noncommutative case, the issue of statistics is closely related to the symmetry of the noncommutative spacetime on which the dynamics is being studied.
If a symmetry acts on a noncommutative spacetime, its coproduct usually has to be twisted in order to make the symmetry action compatible with the  algebraic structure. In the commutative case, particle statistics is superselected, i.e. it is preserved under the action of the symmetry. In the presence of noncommutativity, it is thus natural to demand that the statistics is invariant under the action of the twisted symmetry. This condition leads to a new twisted flip operator, which is compatible with the twisted coproduct of the symmetry group \cite{S3,S4}. The operators projecting to the symmetric and antisymmetric sectors of the Hilbert space are then constructed from the twisted flip operator. While most of the discussion of statistics in the noncommutative setup has been done in the context of the Moyal plane, some
related ideas for $\kappa$-deformed spaces have also appeared recently \cite{S7,S8,S9}.

In this paper we set up a general formalism to describe statistics in $\kappa$-deformed spaces. Our formalism presented here is applicable to a system with an arbitrary symmetry group, which may include Poincare, Lorentz, Euclidean and so on. In particular, we present a general formula for the twisted flip operator that is applicable for any symmetry group, which is constructed in terms of the commutative flip operator, the undeformed coproduct of the symmetry group and the twist element for the $\kappa$-deformed space. In addition, we show that for a special class of realization considered here, the twisted flip operator is independent of the choice of ordering.
This property is valid only for a special class of realizations considered here and is not true for any arbitrary realization. This paper is organized as follows. In Section II we briefly recall some basic facts about $\kappa$ space and discuss a special class of realizations. In Section III we discuss the star product and twist operator and obtain their expression for the specific class of realizations discussed here. In Section IV we introduce the idea of twisted statistics for the $\kappa$ space and obtain a general formula for the twisted flip operator. We also obtain the corresponding expression for the special class of realizations considered here and discuss the ordering independence for this case. Section V concludes the paper with some discussions.

\section{ $\kappa$ Space and its Realizations}
In this Section we briefly review the formalism that leads to a derivation of the Drinfeld twist for the $\kappa$-Minkowski space. Following \cite{paper1} we shall first discuss the $\kappa$-deformed Euclidean space and then analytically continue to the corresponding Minkowski space.

Consider the Lie algebra type relations
\begin{equation} \label{1}
[\hat{x}_{\mu},\hat{x}_{\nu}] = iC_{\mu\nu\lambda}\hat{x}_{\lambda}
= i(a_{\mu}\hat{x}_{\nu}-a_{\nu}\hat{x}_{\mu}),
\end{equation}
defining a noncommutative (NC) space with coordinates
$ \; \hat{x}_{1}, \hat{x}_{2},..., \hat{x}_{n}, \; $, and
where $ a_{1},a_{2},...,a_{n}$ are real constants parametrizing the deformation of the Euclidean space. We choose   $ a_{1}=a_{2}=...=a_{n-1}=0$ and $ a_{n}=a = \frac{1}{\kappa}$  as a very small length scale. Here we use  Latin
indices  for the subspace $(1,2,...,n-1)$ and Greek indices  for
the whole space $(1,2,...,n)$. Then the algebra of
the NC coordinates becomes
\begin{equation} \label{com1}
\lbrack\hat{x}_{i},\hat{x}_{j}\rbrack=0, \qquad
\lbrack\hat{x}_{n},\hat{x}_{i}\rbrack=ia\hat{x}_{i},\qquad
i,j=1,2,...,n-1.
\end{equation}

The NC coordinates $\hat{x}_{\mu}$ admit realizations in terms of ordinary commutative
coordinates  $ x_{1},x_{2},...,x_{n}$ and their derivatives
$\partial_{1},\partial_{2},...,\partial_{n},$  where
$\partial_{\mu}=\frac{\partial}{\partial x_{\mu}}$, which generate the Heisenberg algebra
\begin{equation} \label{2}
 [{\partial}_{\mu}, x_{\nu}] = {\delta}_{\mu \nu}.
\end{equation}
The realizations we want to find can be written in the form
\begin{equation} \label{3}
 \hat{x}_{\mu} = x_{\alpha} {\phi}_{\alpha \mu} ({\partial}_x).
\end{equation}
They will satisfy the condition
\begin{equation} \label{4}
 [{\partial}_{\mu}, \hat{x}_{\nu}] = {\phi}_{\mu \nu}({\partial}_x).
\end{equation}
Here we shall concentrate on the special case of the algebra (\ref{4}),
given  by the relations
\begin{equation}
 [ {\partial}_{i}, {\hat{x}}_{j}] = {\delta}_{ij} \varphi (A);   \; \quad
  [ {\partial}_{i}, {\hat{x}}_{n}] = i a {\partial}_{i} \gamma (A),
\end{equation}
\begin{equation} \label{5}
 [ {\partial}_{n}, {\hat{x}}_{i}] = 0;   \; \quad
  [ {\partial}_{n}, {\hat{x}}_{n}] = 1,
\end{equation}
where $\varphi$ and $\gamma$ are functions of $A=ia\partial_{n}$.
An explicit form of a class of realizations satisfying these conditions is given by
\begin{equation} \label{mapping}
\begin{array}{c}
\hat{x}_{i}=x_{i}\varphi(A),\\
\\
\hat{x}_{n}=x_{n}\psi(A)+iax_{k}\partial_{k}\gamma(A),\\
\end{array}
\end{equation}
where the summation over repeated indices is understood.
The functions $\; \varphi \;$ and $\; \psi \; $ satisfy the conditions
$\varphi(0)=1,\quad  \psi(0)=1$ and they are assumed to be positive.
The function $ \; \gamma \; $ satisfies $\frac{\varphi'}{\varphi}\psi=\gamma-1$,
where $\varphi'=\frac{d\varphi}{dA}$. It is also assumed that the
quantity $\gamma(0)=\varphi'(0)+1$ is finite.
Further conditions on the functions $\varphi$ and $\psi$ are obtained by considering the Lorentz generators in this spacetime. Imposing the conditions that the Lorentz generators be linear in $x$ with an infinite series in $\partial$, leads to two families of realizations given by $ \psi = 1$
and $ \psi = 1 + 2A $. Here we consider only the case $ \psi = 1$, for which the realizations are parametrized by an arbitrary
positive function $\varphi(A),$ \quad $ \varphi(0)=1.$

The coproduct $ \; {\triangle}_{\varphi} \; $
 corresponding to partial derivatives appearing in (\ref{mapping}) is given by
$$
 {\triangle}_{\varphi} ( {\partial}_{n}) = {\partial}_{n} \otimes 1 +
 1 \otimes {\partial}_{n} = {\partial}_{n}^{x} + {\partial}_{n}^{y},
$$
\begin{equation} \label{partialcoproduct}
 {\triangle}_{\varphi} ( {\partial}_{i}) = {\partial}_{i}^{x}
    \frac{\varphi ( A_{x} + A_{y})}{\varphi (A_{x})} + {\partial}_{i}^{y}
   \frac{\varphi ( A_{x} + A_{y})}{\varphi (A_{y})} e^{A_{x}},
\end{equation}
where $ \; A_{x} = ia {\partial}_{n}^{x}  \; $ and similarly for $ \;
A_{y}.$
The last equation can also be written in a more compact way as
\begin{equation} \label{partialcoproductcompact}
 {\triangle}_{\varphi} ( {\partial}_{i})  =  \varphi (A \otimes 1 + 1 \otimes A) \bigg [
   \frac{{\partial}_{i}}{\varphi (A)} \otimes 1 + e^A \otimes
   \frac{{\partial}_{i}}{\varphi (A)} \bigg ].
\end{equation}

There is a well defined correspondence between the realizations defined above and the ordering prescription in the $\kappa$-space \cite{paper1}. To describe it first note that we take the vacuum state to be represented by  $1$, i.e. $\; |0 \rangle \equiv 1, \; $ so that it is annihilated by all derivatives $\; {\partial}_{\mu} $.
We now introduce $\varphi $-ordering $\; :e^{ik \cdot
  {\hat{x}}_{\varphi}}:_{\varphi}   \; $  defined by
the relation
\begin{equation} \label{6}
 :e^{ik \cdot {\hat{x}}_{\varphi}}:_{\varphi} |0 \rangle
 = e^{ik \cdot x}.
\end{equation}
As special cases, the expressions for the left, right and symmetric ordering prescriptions are respectively given by
$$
 :e^{ik_{\alpha} {\hat{x}}_{\alpha}}:_{L} \equiv e^{ik_{n} {\hat{x}}_{n}}
        e^{ik_{i} {\hat{x}}_{i}}  e^{ik_{n} {\hat{x}}_{n} + ik_{i} {\hat{x}}_{i} {\varphi}_{S}(-ak_{n})
 e^{- a k_{n}}},
$$
$$
 :e^{ik_{\alpha} {\hat{x}}_{\alpha}}:_{R} \equiv
        e^{ik_{i} {\hat{x}}_{i}} e^{ik_{n} {\hat{x}}_{n}}  e^{ik_{n} {\hat{x}}_{n} + ik_{i} {\hat{x}}_{i} {\varphi}_{S}(-ak_{n})},
$$
\begin{equation} \label{specialordering}
 :e^{ik_{\alpha} {\hat{x}}_{\alpha}}:_{S} = e^{ik_{\alpha} {\hat{x}}_{\alpha}},
\end{equation}
where
\begin{equation} \label{symmetric}
 {\varphi}_{S}(A) = \frac{A}{e^{A} - 1}, ~~~~~~~~ A = ia {\partial}_{n} \equiv -a k_{n}.
\end{equation}
Corresponding to the realizations of the form (\ref{mapping}), there also exists a family of ordering prescriptions
interpolating between left and right,  parametrized by the function
 $ \; \varphi = e^{- \lambda A}. $ This ordering prescription looks like
\begin{equation} \label{amelino}
 :e^{ik_{\alpha} {\hat{x}}_{\alpha}}:_{\lambda} = e^{i \lambda k_{n} {\hat{x}}_{n}}
            e^{ik_{i} {\hat{x}}_{i}} e^{i (1 - \lambda)k_{n} {\hat{x}}_{n}},
\end{equation}
 For $ \; \lambda = 0, \; $ we obtain a right ordering, while for $
  \; \lambda = 1, \; $ a left ordering emerges. For
  $ \; \lambda = \frac{1}{2}, \; $ the same factor appears on the left
  and right hand side symmetrically \cite{fedele}. In this sense it can be named as
  symmetric ordering, but it should be kept in mind that it is
  completely different from the totally symmetric Weyl ordering.

More generally, corresponding to the $\varphi$ ordering given by
(\ref{mapping}), the ordering prescription can be written as
\begin{equation} \label{ordering}
 :e^{ik_{\alpha} {\hat{x}}_{\alpha}}:_{\varphi}   e^{ik_{n} {\hat{x}}_{n} + ik_{i} {\hat{x}}_{i} \frac{{\varphi}_{S}(-ak_{n})}
 {\varphi(-ak_{n})}},
\end{equation}
which contains all the special cases described above.
This gives the direct correspondence between the various realizations and the corresponding ordering prescriptions.


\section{ Star Product and Twist in $\kappa$ Space }

For any $\varphi$ ordering, a star product can be defined in the $\kappa$ space. This star product can be expressed in terms of the corresponding twist element ${\mathcal{F}}_{\varphi}$ as
\begin{equation} \label{star1}
f \; {\star}_{\varphi} \; g  =  m_{0} ( {\mathcal{F}}_{\varphi} f \otimes g) =  m_{\varphi}(f
  \otimes g), \quad m_{\varphi} = m_{0}{\mathcal{F}}_{\varphi}
\end{equation}
where $ \; m_0 \; $ is the usual pointwise multiplication map in the commutative algebra of smooth functions ${\mathcal{A}_0}$, namely, $ \; m_0 (f \otimes g) = f g. $ and $f,g \in {\mathcal{A}_0}$.

The star product of two functions $f$ and $g$ in the $\varphi$ realization is given by
\begin{equation} \label{fstarg}
(f \; \star_{\varphi} \; g)(x)  =   \lim_{\substack{u \rightarrow x  \\ t \rightarrow x }}
 m_0 \left ( e^{x_{\alpha} ( \triangle_\varphi - {\triangle}_{0}) {\partial}_{\alpha} }
    f(u) \otimes g(t) \right ),
\end{equation}
where ${\triangle}_{\varphi}$ is given in (\ref{partialcoproductcompact}) and
\be \label{untwist}
{\triangle}_{0} (\partial ) = \partial \otimes 1 + 1 \otimes \partial
\ee
is untwisted coproduct. The corresponding
twist element is given by
\begin{equation} \label{tphi}
 {\mathcal{F}}_{\varphi}
 = e^{x_{\alpha} ({\Delta}_{\varphi} - {\Delta}_{0}) {\partial}_{\alpha}},
\end{equation}
where ${\triangle}_{\varphi}$ satisfies the relation
\begin{equation} \label{cophi}
 {\triangle}_{\varphi} = {{\mathcal{F}}_{\varphi}}^{-1}  {\triangle}_{0}
            {\mathcal{F}}_{\varphi}.
\end{equation}

Using (\ref{fstarg}) and (\ref{partialcoproduct}), the star product in the $\kappa$-space can be written as
\begin{equation} \label{star}
(f \; {\star}_{\varphi} \; g)(x)
  =  \lim_{\substack{u \rightarrow x  \\ t \rightarrow x }}
 e^{x_{j} {\partial}_{j}^{u} \left (
  \frac{ \varphi (A_{u} + A_{t}) }{\varphi (A_{u})} - 1  \right )
     + x_{j} {\partial}_{j}^{t} \left (
        \frac{ \varphi (A_{u} + A_{t}) }{\varphi (A_{t})} e^{A_{u}} - 1 \right )}
    f(u) g(t),
\end{equation}
where we use the notation $ \; A_{x}=ia\frac{\partial}{\partial x_{n}}, \; $ etc.
The corresponding twist operator (\ref{tphi}) can be written as
\bea \label{twist}
{\mathcal{F}}_{\varphi}
 &=& e^{ N_x \ln \frac{\varphi (A_x + A_y )}{\varphi (A_x)}
       + N_y (A_x +  \ln \frac{\varphi (A_x + A_y)}{\varphi (A_y)})} \nonumber \\
&\equiv&  e^{ (N \otimes 1) \ln \frac{\varphi (A \otimes 1 + 1 \otimes A)}{\varphi (A \otimes 1)}
       + (1 \otimes N) (A \otimes 1 +  \ln \frac{\varphi (A \otimes 1 + 1 \otimes
 A)}{\varphi (1 \otimes A)})},
\eea
 where
$N_x = x_i \partial / \partial x_i $ and similarly for $N_y$. In
writing the above expression of the twist operator we have used
the identity \cite{paper1} (see eqs (A.16),(A.17))
\begin{equation} \label{toperator}
   \lim_{u \rightarrow x} e^{x_{i} {\partial}_{i}^{u} \left [
     \Phi(A_u,A_y) - 1 \right ] }
 = e^{x_{i} {\partial}_{i}^x \ln \Phi(A_x,A_y)},
\end{equation}
valid for arbitrary function $\Phi(A_u,A_y)$.

We point out that the star product and the twist element depends
explicitly on the choice of the ordering. For example, the twist
operator corresponding to the left ordering is given by
\begin{equation} \label{tl}
 {\mathcal{F}}_{L}
 = e^{- N_{x} A_{y}} = e^ {-N \otimes A},
\end{equation}
while that corresponding to the right ordering is
\begin{equation} \label{tr}
  {\mathcal{F}}_{R}
 = e^{ A_{x} N_{y}} = e^{A \otimes N}.
\end{equation}
For $\phi(A)=e^{-cA}$ where $c \in R$, we obtain a simple interpolation between right ordering ($c = 0$) and left ordering ($c = 1$), with the twist operator given by
\begin{equation} \label{interpol}
{\mathcal{F}}_c = e^{-cN \otimes A + (1-c)A \otimes N}.
\end{equation}
Using  $\triangle N = N \otimes 1 + 1 \otimes N$, $\triangle A = A \otimes 1 + 1 \otimes A$, and $[N,A]=0$, it is easy to verify that the above class of twist operators ${\mathcal{F}}_c$ satisfies the cocycle condition
\begin{equation}
({\mathcal{F}}_c \otimes 1)(\triangle \otimes 1){\mathcal{F}}_c = 
(1 \otimes {\mathcal{F}}_c)(1 \otimes \triangle){\mathcal{F}}_c,
\end{equation}
for all $c \in R$.

The above results can be continued to the Minkowski space by redefining the momenta as
\begin{equation} \label{impuls}
   P_{0} = i {\partial}_{0},  \;\;\;\;  P_{i} = - i {\partial}_{i},
   \;\;\;\; i = 1,2,...,N-1,
\end{equation}
which satisfy
\begin{equation} \label{impuls0}
  [P_{\mu}, x_{\nu}] = - i {\eta}_{\mu \nu},  \quad   \mu, \nu = 0,1,..., N-1,
\end{equation}
$ \; {\eta}_{\mu \nu} = diag(-1,+1,+1,...,+1). \;$
The algebra of the variables $ \; x_{\mu}, {\partial}_{\nu}  \; $
remains unaltered,
\begin{equation} \label{impuls1}
  [{\partial}_{\mu}, x_{\nu}] = {\delta}_{\mu \nu},  \quad   \mu, \nu = 0,1,..., N-1.
\end{equation}
Accordingly, the commutation relation describing $ \; \kappa$-deformed Minkowski space takes the form
\begin{equation} \label{com2}
\lbrack\hat{x}_{i},\hat{x}_{j}\rbrack=0, \qquad
\lbrack\hat{x}_{0},\hat{x}_{i}\rbrack=ia\hat{x}_{i},\qquad
i,j=1,2,...,N-1,
\end{equation}
where $x_n = i x_0$ and the old deformation parameter $a_n$ is related to the new one $a_0$ by $ a_n = i a_0$. For simplicity, we remove the index 0 of it, but in all subsequent
 consideration one should keep in
 mind that it is the time component of the $ \; N$-vector in Minkowski space
and that it is real. The definition of the star product also remains unchanged, provided in (\ref{star}) we replace $ \; A_{x} \; $ by  $ \; A_{x} = ia \partial_{0}^{x} \; $.

In terms of the $\kappa$-Minkowski space variables, as defined above, we can write the coproduct and the twist operator in the momentum space as well. The coproduct $\; K_{\varphi} (p,q)  \;$ in momentum space is given by
\begin{equation} \label{momcop}
\begin{array}{c}
{\bigg [K_{\varphi} (p,q) \bigg ]}_{\mu} = -i {\triangle}_{\varphi} ( {\partial}_{\mu}),\\
\\
K_{\varphi} (p,q) x = -(k_0 + q_0) x_0 + \varphi (ak_0 + aq_0)
   \bigg [ \frac{k_i  x_i}{\varphi (ak_0)} + \frac{e^{ak_0}}{\varphi (aq_0)}
 q_i  x_i \bigg ].\\
\end{array}
\end{equation}
Let ${\mathcal{F}}$ denote the corresponding twist operator in the momentum space. Its defining relation is
$$
{\mathcal{F}} f(x) g(y) \equiv {\mathcal{F}} (x, p^{x}, y, p^{y})  f(x) g(y)
 = {\mathcal{F}} \int d^{4}k e^{ikx} {\tilde{f}} (k) \int d^{4}q e^{iqy} {\tilde{g}} (q)
$$
$$
= \int d^{4}k d^{4}q \bigg ( {\mathcal{F}} e^{ikx} e^{iqy} \bigg )
{\tilde{f}} (k) {\tilde{g}} (q)   \int d^{4}k d^{4}q  e^{ikx} e^{iqy} {\mathcal{F}} (i
  \frac{\partial}{\partial k}, k, i \frac{\partial}{\partial q}, q) {\tilde{f}} (k) {\tilde{g}} (q).
$$
From this we obtain
\be \label{momtwist}
 {\mathcal{F}} {\tilde{f}} (k) \otimes {\tilde{g}} (q)
 = {\mathcal{F}} (i \frac{\partial}{\partial k}, k, i
 \frac{\partial}{\partial q}, q) {\tilde{f}} (k) {\tilde{g}} (q).
\ee


\section{ Twisted Statistics in $\kappa$ Space}

Consider a system of particles or fields described by an action principle. Very often the action describing such a system remains invariant when acted upon by a suitable group, which is called the symmetry group of the system. For example, the symmetry group could be the Poincare group, Lorentz group, Euclidean group and so on. In what follows we assume that the system under consideration is invariant under a symmetry group denoted by $G$.

In the commutative case, the identical particles or fields usually satisfy either bosonic or fermionic statistics. On a two particle Hilbert space, the bosonic or fermionic statistics is implemented by restricting to the symmetrized and antisymmetrized sectors respectively.
Normally, the two subsectors of the Hilbert space corresponding to bosonic or fermionic statistics are individually preserved under the action of the symmetry group $G$. This is to say that the statistics is superselected.

Let $\Lambda$ be an element of $G$ which acts with some representation $D$. In the commutative case with $a=0$, the action of $G$ on the two particle Hilbert space is described by the coproduct $\Delta_0 $ where
$$
 \Delta_0 ~ :~ \Lambda \longrightarrow \Lambda \otimes \Lambda , ~~~~
$$
\begin{equation}
f \otimes g \longrightarrow  (D \otimes D) \Delta_0(\Lambda) f \otimes g .
\end{equation}
The coproduct $\Delta_0 (\Lambda)$ is compatible with the usual pointwise multiplication map $m_0$, i.e. it satisfies the condition
\begin{equation}
 m_0 \, \left ((D \otimes D) \Delta_0(\Lambda) f \otimes g \right )
= D(\Lambda) \, m_0  (f \otimes g).
\end{equation}


Statistics of identical particles is usually defined in terms of a flip operator $\tau_0$. On an element $f \otimes g \in {\mathcal{A}_0} \otimes  {\mathcal{A}_0}$, it has the action
\begin{equation} \label{t0}
 {\tau}_{0}(f \otimes g) = g \otimes f.
\end{equation}
Symmetrization or antisymmetrization on a two particle Hilbert space is carried out by the operators $\frac{1}{2} ( 1 \pm \tau_0 )$.
In the commutative case, the flip operator $\tau_0$ commutes with the coproduct $\Delta_0 (\Lambda) $ of the symmetry group $G$, i.e.
\begin{equation}
[{\Delta}_{0} (\Lambda), {\tau}_{0}] = 0.
\end{equation}
Physically this means that the process of symmetrization or antisymmetrization is frame independent. The statistics thus remains invariant under the action of the symmetry group G.

In the noncommutative case, when $ a = \frac{1}{\kappa} \neq 0$, the multiplication map in the algebra is defined by the star product, denoted by $m_\varphi$ in the case of a given $\varphi$ realization . In this case, the coproduct $\Delta_0$ of the symmetry group $G$ is no longer compatible with $m_\varphi$. This implies that $G$ cannot be an automorphism of the algebra (\ref{com1}) with the coproduct $\Delta_0$. We can define a twisted coproduct for the symmetry group $G$, denoted by
$\Delta_\varphi$ where
\begin{equation} \label{tcop}
 {\Delta}_{\varphi}(\Lambda) = {{\mathcal{F}}_{\varphi}}^{-1}  {\Delta}_{0} (\Lambda) {\mathcal{F}}_{\varphi},
\end{equation}
and ${\mathcal{F}}_{\varphi}$ is the twist element. This new twisted coproduct
is compatible with the multiplication map $m_\varphi$, i.e. it satisfies the equation
\begin{equation}
 m_\varphi \, \left ((D \otimes D) \Delta_\varphi(\Lambda) f \otimes g \right )
= D(\Lambda) \, m_\varphi  (f \otimes g).
\end{equation}
Henceforth we work with the twisted coproduct $\Delta_\varphi$. The symmetry group $G$ with the twisted coproduct (\ref{tcop}) is an automorphism of the algebra (\ref{com1}) and is often called the twisted symmetry group.

In the presence of $\kappa$ deformation, it is also desirable that the statistics be left unchanged under the action of the twisted symmetry group. This would ensure that even in the presence of the $\kappa$ deformation, the symmetrization procedure would be left invariant under the action of the twisted symmetry group and that the twisted statistics remains superselected. However, it turns out that the flip operator $\tau_0$ does not commute with the twisted coproduct, i.e.
\begin{equation}
[\Delta_\varphi (\Lambda), \tau_0 ] \neq 0.
\end{equation}
This means that the symmetrization or antisymmetrization carried out by $\tau_0$ would not be preserved by the action of the twisted symmetry group. We must therefore seek a new flip operator, called the twisted flip operator $\tau_\varphi$ which would satisfy the condition
\be\label{cond}
[\Delta_\varphi (\Lambda), \tau_\varphi ] = 0.
\ee
Such a twisted flip operator can be defined and is given by
\be \label{tflip}
 \tau_\varphi = {\mathcal{F}}_\varphi^{-1} \tau_0 {\mathcal{F}}_\varphi,
\ee
where ${\mathcal{F}}_\varphi$ is the twist element in (\ref{tphi}). The new flip operator $\tau_\varphi$ commutes with the twisted coproduct of the symmetry group $G$, i.e. it satisfies the condition (\ref{cond}). Having defined the twisted flip operator, the symmetrization and antisymmetrization can now be carried out with the projection operators $\frac{1}{2} (1 \pm \tau_\varphi )$ respectively. This would ensure that the twisted statistics of a two particle state in the $\kappa$ space would remain unchanged under the action of the twisted symmetry group $G$.

The expression of the twisted flip operator $\tau_\varphi$ in (\ref{tflip}) is valid for an arbitrary $\varphi$ realization of the $\kappa$-deformed space and for a general symmetry group $G$ for the dynamics. We now restrict our attention to the special cases of the $\varphi$ realizations given in (\ref{mapping}). For this particular special class of realizations, the twist element is given in (\ref{twist}). Using (\ref{twist}) and (\ref{tflip}), we obtain an explicit expression for the twisted flip operator $\tau_\varphi$ for the class (\ref{mapping}) of realizations as
\be \label{tflip1}
\tau_\varphi = e^{i(x_iP_i \otimes A - A \otimes x_iP_i)}\tau_0.
\ee
Note that as $A$ is directly proportional to the deformation parameter $a$,
the twisted flip operator $\tau_\varphi$ goes over smoothly to the untwisted flip operator $\tau_0$ as the deformation parameter $a \rightarrow 0$.
It may be noted that although the twist element ${\mathcal{F}}_\varphi$ depends explicitly on the choice of the $\phi$ realization, it is a remarkable fact that the twisted flip operator $\tau_\varphi$ as obtained in (\ref{tflip1}) is independent of the class of realizations (\ref{mapping}) and the corresponding orderings. This conclusion follows naturally from the properties of the twist element (\ref{twist}) and the twisted flip operator (\ref{tflip1}) in the $\kappa$ space. By construction, any symmetrization or antisymmetrization of two particle states carried out using the flip operator $\tau_\varphi$ would be preserved under the action of the Poincare group with a twisted coproduct. Thus, the twisted flip operator obtained above is
a good candidate to describe particle statistics in the $\kappa$ space. We however point out that the $\varphi$ independence of twisted flip operator is valid only for the class of realization (\ref{mapping}) considered in this paper and will not hold in the case of an arbitrary realization.

In the paper by Daszkiewicz, Lukierski and Woronowicz \cite{S9}, the quantized scalar field on
$\kappa$-spacetime was considered. Calculations performed in \cite{S9},
especially the coproduct, correspond to our realization $\phi(A)=e^{-A/2}$.
In classical limit their results are equivalent to our statistics flip
operator equation (\ref{tflip1}). Arzano and Marciano \cite{S7} construct their twist operator
to be consistent with kappa Poincare algebra, whereas in our construction
we have an arbitrary symmetry algebra $G$. In the limit kappa going to infinity both lead to
commutative theory. It may also be noted that in our approach the Lorentz algebra is undeformed and corresponding coalgebra is $\kappa$ deformed. The effect of this will be reflected in correlation functions. The quantization of scalar field on $\kappa$-Minkowski space along within our framework and its relation to \cite{S7,S9} is presently under investigation.

We end this Section with some remarks about the CPT symmetry and spin-statistics theorem for the $\kappa$-deformed space. For the Moyal plane with space-time noncommutativity these issues have been discussed in \cite{p1}. Let us consider a system described by the $\kappa$-deformed Minkowski space (\ref{com2}). The parity operator $P$, which is linear, can be defined as
\be \label{parity}
P ~ : ~  \hat{x}_{0} \rightarrow \hat{x}_{0}, ~~ \hat{x}_{i} \rightarrow - \hat{x}_{i}, ~~ ia \rightarrow ia.
\ee
This is an automorphism of the algebra (\ref{com2}). Hence, unlike in the Moyal case \cite{p1}, parity is a symmetry of the $\kappa$-deformed Minkowski space.

The time reversal operator $T$, which is anti-linear, is defined by
\be \label{time}
T ~ : ~  \hat{x}_{0} \rightarrow -\hat{x}_{0}, ~~ \hat{x}_{i} \rightarrow  \hat{x}_{i}, ~~ ia \rightarrow -ia.
\ee
Thus $T$ and is also an automorphism of the algebra (\ref{com2}) and the same holds for $PT$. In case the charge conjugation is defined and is an automorphism, then $CPT$ can be realized as an automorphism on the $\kappa$-deformed Minkowski space.

The star product for the $\kappa$-deformed Minkowski space, just as in the Moyal case, is a nonlocal quantity. The lack of locality leads to a failure of causality \cite{p2}, and the spin-statistics theorem also fails to be valid. In particular, the Pauli exclusion principle is not expected to be valid for theories based on the algebra (\ref{com2}), and bounds on the deformation parameter $a$ can be obtained from the bounds on the validity of the Pauli exclusion principle \cite{S3,S4}.


\section{ Conclusion}

In this paper we have discussed the issue of statistics for a system of identical particles or a field theory in $\kappa$ deformed space, where the symmetry group of the system under consideration is described by a general group $G$. Due to the noncommutative structure of the $\kappa$ deformed space, the coproduct of the symmetry group $G$ has to be twisted. The flip operator describing the statistics in the commutative case no longer commutes with the twisted coproduct of the symmetry group. As a result, the symmetrization or antisymmetrization carried out by the projection operators obtained from the commutative flip operator are no longer preserved under the action of the symmetry group. This leads us to seek a new twisted flip operator that would commute with the twisted coproduct of $G$, and we have presented a general expression of the twisted flip operator for an arbitrary realization of the $\kappa$ deformed space and for a general symmetry group $G$. The twisted stat
 istics obtained from the twisted flip operator remains superselected.

A $\kappa$ deformed space admits many different realizations. We have restricted our attention to a particular class in this paper. It has been shown that within this class, the twisted flip operator is independent of the choice of the realization and the corresponding ordering. The projection operators corresponding to the symmetric and antisymmetric sectors of the Hilbert space also share this property for this special class. This property is not valid for any arbitrary realization of the $\kappa$ deformed space, not belonging to the class described here.

The analysis here provides a general framework for considering statistics in $\kappa$ deformed spaces which would be relevant for the analysis of field theories in such spaces. The implications of the twisted statistics for quantum field theories on $\kappa$ deformed spaces is currently under investigation.


\vskip 1cm

\noindent
{\bf Acknowledgment}\\
The authors would like to thank A. Samsarov and Z. Skoda for discussions.
This work was supported by the Ministry of Science and Technology of the Republic of Croatia under contract No. 098-0000000-2865. This work was done within the framework of
the Indo-Croatian Joint Programme of Cooperation in Science and Technology
sponsored by the Department of Science and Technology, India (DST/INT/CROATIA/P-4/05), and
the Ministry of Science, Education and Sports, Republic of Croatia.

\end{document}